
\overfullrule=0pt
\parindent=24pt
\baselineskip=18pt
\magnification=1200

\def\dlsim{\vbox{\hbox{\raise0.9mm\hbox{$<$}}
	\kern-18pt\hbox{\lower0.90mm\hbox{$\sim$}}}}
\def\dgsim{\vbox{\hbox{\raise0.9mm\hbox{$>$}}
	\kern-18pt\hbox{\lower0.90mm\hbox{$\sim$}}}}
\def\lsim{\vbox{\hbox{\raise0.9mm\hbox{$<$}}
	\kern-18pt\hbox{\lower0.90mm\hbox{$\sim$}}}}
\def\gsim{\vbox{\hbox{\raise0.9mm\hbox{$>$}}
	\kern-18pt\hbox{\lower0.90mm\hbox{$\sim$}}}}

\def\rl{\rightline}
\def\cl{\centerline}

\def\e6{{\rm E}_6}
\def\su#1{{\rm SU(#1)}}

\def\bsg{b\rightarrow s\gamma}
\def\NPB#1#2#3{Nucl. Phys. {\bf B#1}, #2 (19#3)}
\def\PLB#1#2#3{Phys. Lett. {\bf#1B}, #2 (19#3)}
\def\PRD#1#2#3{Phys. Rev. D {\bf#1}, #2 (19#3)}
\def\PRL#1#2#3{Phys. Rev. Lett. {\bf#1}, #2 (19#3)}
\def\PRT#1#2#3{Phys. Rep. {\bf#1C}, #2 (19#3)}

\def\MPLA#1#2#3{Mod. Phys. Lett. {\bf A #1}, #2 (19#3)}

\def\ibid#1#2#3{{{\it ibid}, {#1}, #2 (19#3)}}


\def\bsg{b\rightarrow s\gamma}
\def\BR{{\rm BR}}
{\nopagenumbers
	\rl{CTP-TAMU-03/94}
	\rl{NUB-TH.7316/94}
	\rl{CERN-TH.3092/94}
	\rl{March, 1994}
\medskip
	\cl{\bf Constraints on the minimal supergravity model}
	\cl{\bf from the $\bsg$ decay}
\medskip
	\cl{Jizhi Wu\quad and\quad Richard Arnowitt}
	\cl{Department of Physics, Texas A\&M University}
	\cl{College Station, Texas 77843}
	\cl{and}
	\cl{Pran Nath}
	\cl{Theoretical Physics Division, CERN}
	\cl{CH-1211 Geneva 23}
	\cl{and}
	\cl{Department of Physics, Northeastern University}
	\cl{Boston, Massachusetts 02115\footnote{$^*$}{Permanent address}}
\bigskip
	\centerline{\bf Abstract}
\medskip
	The constraints on the minimal supergravity model from the
	$\bsg$ decay are studied.
	A large domain in the parameter space for the model satisfies the
	CLEO bound, $\BR(\bsg)<5.4\times10^{-4}$.
	However, the allowed domain is expected to diminish significantly
	with an improved bound on this decay.
	The dependence of the $\bsg$ branching ratio on various parameters
	is studied in detail.
	It is found that, for $A_t<0$ and the top quark mass within the
	vicinity of the center of the CDF value, $m_t^{pole}=174\pm$17~GeV,
	there exists only a small allowed domain because the light stop is
	tachyonic for most of the parameter space.
	A similar phenomenon exists for a lighter top and $A_t$ negative
	when the GUT coupling constant is slightly reduced.
	For $A_t>0$, however, the branching ratio is much less sensitive
	to small changes in $m_t$, and $\alpha_G$.
\medskip
	PACS numbers: 12.10Dm, 11.30.Pb, 14.65.Fy, 14.40.Nd
\vfill
\eject}
\pageno=1

	The extensive analyses of the high precision LEP data in the last
    	few years have indicated that the idea of grand unification is
    	only valid when combined with supersymmetry~[1].
    	One of most promising and most studied models is the minimal
    	supergravity model (MSGM)~[2].
	Supersymmetry (SUSY) is naturally and softly broken by a hidden sector.
\def\mgut{M_{GUT}}
    	In addition to Yukawa couplings, the gauge coupling constant,
	$\alpha_G$, and the unification scale, $\mgut\simeq10^{16}$~GeV,
	there are only five free parameters in this model: the four soft
	breaking terms [the universal scalar mass, $m_0$, the universal
	gaugino mass, $m_{1/2}$, the cubic scalar coupling, $A_0$, and the
	quadratic scalar coupling, $B_0$], and a supersymmetric Higgs mixing,
	$\mu_0$.
	The common approach to constraining the GUT model is to utilize the
    	renormalization group equations (RGE) to make contact with physics
	at the electroweak scale, $M_{EW}$~[3].
    	Remarkably, the evolution of RGEs from $\mgut$ to $M_{EW}$ produces
	a Higgs potential with a negative $m_{H_2}^2$ if the top quark is
	heavy, signaling a spontaneous breaking of the electroweak gauge
	symmetry.
    	As a consequence of this radiative breaking, two constraints arise
	which relate the GUT parameters to the electroweak parameters.
    	One may then eliminate two GUT parameters, $B_0$, and $\mu_0$, in
	favor of two electroweak parameters: the Higgs VEV ratio
    	$\tan\beta\equiv v_2/v_1$, and the $Z$ boson mass, $M_Z$.
    	Therefore, the low-energy physics depends only on four parameters
\def\tanbeta{\tan\beta}
$$
    m_0, ~ m_{\tilde g}, ~ A_t, ~ \tanbeta,
\eqno(1)
$$
	and the sign of $\mu$ [since the renormalization group
	equations determine only $\mu^2$].
    	Here, we have replaced $m_{1/2}$ by the gluino mass,
    	$m_{\tilde g}$, and $A_0$ by $A_t$, their values at the electroweak
	scale, $M_{EW}$.
    	In most analyses, only the masses of the third generation of
	leptons and quarks are important.
    	For small ${\rm tan} \beta$, one needs to retain only the top
	quark mass.
    	Hence, the four free parameters in eq.~(1), plus the top quark
	mass, $m_t$, suffice to parametrize the MSGM.

	The constraints on the parameter space of the MSGM may be classified
	into three major categories.
        First, various theoretical considerations put stringent constraints on
	the parameter space.
        For example, the color $\su3_C$ group should remain unbroken when
	discussing the radiative breaking; $\mu^2$ should also stay positive
	to guarantee this breaking; all scalar particles must be non-tachyonic;
	and the allowed parameter space should be such that theory remains in
	the perturbative domain.
        Some of these issues will be discussed in a separate paper~[4].
        Second, cosmological considerations and the proton stability
        also strongly constrain the model~[3].
	Thus, for \su5-type models, proton stability requires that
	$\tanbeta$ should not be too large, i.e., $\tanbeta\lsim10$~[5].
        There still exists a large domain in the parameter space which
	satisfies both the proton decay and the relic density bounds~[6].
    	Third, there exist a vast amount of data from the electroweak physics
	in the low-energy domain.
	It is thus very interesting to use these data to constrain the MSGM.
	One of the interesting processes is the $\bsg$ decay.
	This decay is very sensitive to the structure of fundamental
        interactions at $M_{EW}$, because its rate is of order $G_F^2\alpha$,
	while most other FCNC processes are of order $G_F^2\alpha^2$.
	We shall study the constraint coming from the $\bsg$ decay in the MSGM
	in this paper.
	The combined constraints, from $\bsg$,proton decay,and relic density
	will be discussed elsewhere [7].

	The recent CLEO II experiment gives the following
\def\BKG{B\rightarrow K(892)^*\gamma}
     	measurement for the branching ratio of the exclusive $\BKG$ decay~[8],
    $\BR(\BKG) = (4.5 \pm 1.5 \pm 0.9) \times 10^{-5}$,
\def\BXSG{B\rightarrow X_s\gamma}
	and an upper bound on the inclusive $\BXSG$ decay,
    $\BR(\BXSG) < 5.4 \times 10^{-4}$.
	Much work has been devoted to the determination of the ratio
	$\BR(\BKG)/\BR(\BXSG)$ from QCD calculations~[9-11].
	An accurate calculation is of interest since the inclusive $B$ decay
	is not as easy to study experimentally as the exclusive $B$ decay.
	The earlier calculations~[9] made use of the constituent-quark
	model and the 2-point QCD sum rules, and the results for this
	ratio ranged from 0.05 to 0.40.
	More recently, there have appeared calculations based on some new
	developments in QCD.
	The method based on the 3-point QCD sum rules predicts a value of
	$0.17\pm0.05$~[10].
        The method that incorporates the chiral symmetry for the light
        degrees of freedom and the heavy quark spin-flavor symmetries
 	for the heavy quarks gives a value of $0.09\pm0.03$ for this
	ratio~[11].
	Although these results are in accord with the CLEO II measurement,
	more study is needed to fully utilize the CLEO result.
	Because the $b$ quark mass is much greater than the QCD scale, the
	dominant contributions to the inclusive $B$ decay are from the
	short-range interactions.
	Barring a large interference between the long- and short-range
	contributions, one may model the inclusive $B$ transitions, such
\def\BXC{B\rightarrow X_ce{\bar\nu}_e}
	as $\BXSG$ and $\BXC$, by the decay of a free $b$ quark to a free
        light quark, such as a free $s$ or $c$ quark.
	In fact, a calculation based on the heavy quark effective theory
	shows that $\BR(\BXSG)$ agrees with the free quark results, $\BR(\bsg)$
	up to corrections of order $1/m_b^2$~[12].
	Thus, the bound on $\BR(\BXSG)$ transfers to an upper bound,
	$\BR(\bsg)<5.4\times10^{-4}$.
	It is a common practice to use the ratio defined as
\def\bcenu{b\rightarrow ce{\bar\nu}_e}
$$
   R={\BR(\bsg)\over{\BR(\bcenu)}}
    \simeq {\BR(\BXSG)\over{\BR(\BXC)}}.
\eqno(2)
$$
	to constrain various models, utilizing the well-determined value
        of $(10.7 \pm 0.5)\%$ for $\BR(\BXC)$.
	The advantage of using $R$, instead of $\BR(\bsg)$, is that the latter
	is dependent upon $m_b^5$ and certain elements of the Kobayashi-Maskawa
	matrix, while the former only depends on $z=m_c/m_b$, the ratio between
	the $c$ and $b$	quark masses, which is much better determined than
	both masses, i.e., $z=0.316 \pm 0.013$~[13].

	The ratio $R$ defined in eq.~(2) has been calculated as~[14]
\def\c#1w{C_{#1}(M_Z)}
$$
    R={6\alpha\over\pi}
      {[\eta^{16/23}\c7w+{8\over3}(\eta^{14/23}-\eta^{16/23}\c8w)
        +\c2w]^2\over{I(z)[1-{2\over3\pi}{\alpha_s(M_Z)\over\eta}f(z)]}}
\eqno(3)
$$
	where $\eta=\alpha_s(M_Z)/\alpha_s(m_b)=0.548$.
	Here, $I(z)=1-8z^2+8z^6-z^8-24z^4\ln z$ is the phase-space
	factor, and $f(z)=2.41$, a QCD correction factor, for the
	semileptonic process, $\bcenu$.
	$\c7w$ and $\c8w$ are the coefficients of the photonic and gluonic
	penguin operators for the $bs$ transition at the electroweak scale.
	These coefficients are model dependent and sensitive to the underlying
	fundamental interactions at $M_Z$.
	For the standard model (SM), only the penguin diagram induced by the
	$W$-$t$ loop contributes to $\c7w$ and $\c8w$, whereas, for the MSGM,
	many susy particles contribute.
	$\c2w$ is a coefficient coming from a mixing between the photonic
	penguin operator and many four-quark operators present at $M_Z$.
	The form and number of these four-quark operators differ depending on
	the model.
	Fortunately, for the MSGM, they are the same as those for the SM~[15].
	The calculation of $\c2w$ is an involved procedure and a number
	of evaluations exist~[14,16,17].
	We shall use the results of Ref.~[17], which takes into
	account of the full-leading-order logarithmic contributions,
	i.e., $\c2w=\sum_{i=1}^8a_i\eta^{b_i}=-0.1795$, where $a_i$ satisfy
	$\sum_ia_i=0$ [since at the electroweak scale, only the photonic
	penguin diagram has a contribution to the $\bsg$ decay].
	The numerical values for $a_i$ and $b_i$ have been a matter of
	debate, in the sense that different calculations give different
	results.
	Nevertheless, the over all effect on $\c2w$ is insignificant, the
	difference being only about $1\%$.
	A comment on the accuracy of eq.~(3), however, is in order.
	There exist several uncertainties in using this equation to calculate
	$\BR(\bsg)$.
	For instance, eq.~(2) is based on the spectator quark model.
	The error in determining the strong interaction constant,
\def\alphasw{\alpha_s(M_Z)}
	$\alphasw$, is still large.
	The most significant uncertainties come from the absence of a complete
	evaluation of the next-to-leading short-distance QCD corrections to
	$\BR(\bsg)$, causing about a $25\%$ inaccuracy.
	To better determine the theoretical predictions for $\BR(\BXSG)$,
	it would be necessary to calculate certain three-loop mixings and
	two-loop penguin diagrams~[18].


	In the rest of this paper, we will concentrate on the MSGM
	predictions for $\BR(\bsg)$.
	As mentioned above, this model contains many particles not present
	in the standard model.
	Thus, besides the $W$ boson contributions, there exist the penguin
	diagrams induced by the charged Higgs bosons, the chargino-squarks,
	the gluino and the neutralinos.
	The coefficients $\c7w$ and $\c8w$ were  calculated for the minimal
	supersymmetric model (MSSM) in Ref.~[19].
	(We have rederived these coefficients confirming their results.)
	It is found that the gluino and neutralino contributions are small
	compared to other sources.
	We will hence ignore their contributions below.
	A large contribution from the charged Higgs boson in the MSSM was found
	in Ref.~[20] and these authors thus concluded that a slight improvement
	in the experimental bound on $\BR(\bsg)$ will exclude the search for
	the charged Higgs boson via the $t\rightarrow bH^+$ decay channel.
	However, these papers did not consider the chargino-squark penguin
	diagrams.
	As it turned out, although the charged Higgs boson enhances the
	standard model amplitude, the chargino-squark loops may contribute to
	the amplitude constructively or destructively, depending on the
	parameters chosen.
	In fact, as shown in Ref.~[21], in the exact supersymmetric limit,
	the coefficients for $bs\gamma$ and $bsg$ transition operators,
	$\c7w$ and $\c8w$, vanish exactly.
	Other papers on the $\bsg$ decay in SUSY models can be found in [22].
	We will follow the notations of Ref.~[21], and assume that the first
	two generations of squarks are degenerate in mass.
	We then expect the contributions from these degenerate squarks to
	$\BR(\bsg)$ to be small, since their masses are proportional to
	$m_0\lsim1$ TeV.
\def\st#1{{\tilde t}_{#1}}
	On the other hand, the scalar top squarks, $\st1$ and $\st2$, are badly
	split in their masses, due to the large top mass, implied by the recent
	CDF data~[23].
	The stop mass matrix is given by
$$
   \left(\matrix{m^2_{\st{L}}&m_t(A_t+\mu\cot\beta)\cr
		 m_t(A_t+\mu\cot\beta)&m^2_{\st{R}}\cr}
	\right),
\eqno(4)
$$
	where $m^2_{\st{L}}$ and $m^2_{\st{R}}$ are given in Ref.~[24].
	Thus, the light stop mass is,
$$
    m^2_{\st1}={1\over2}\left(m^2_{\st{L}}+m^2_{\st{R}}-
	       \sqrt{(m^2_{\st{L}}-m^2_{\st{R}})^2+4m^2_t(A_t+\mu\cot\beta)^2}
	\right).
\eqno(5)
$$
	One can demonstrate that, for a large portion of the parameter space,
	the light stop ${(\rm mass)^2}$ may turn negative, signalling either
	the breaking of the color $\su3_C$ group or the existence of tachyons
	in the theory.
	The requirement that $m^2_{\st1}$ be positive is very stringent and
	could eliminate a considerably large domain in the parameter space~[4].

	Our strategy is to use the one-loop RGEs to calculate the mass spectrum
	of the model relevant to $\bsg$ [all the up-type squarks, the
	charginos, and the charged Higgs bosons].
	We then use this spectrum to evaluate the ratio $R$ via eq.~(3) and
	eq.~(2) to obtain $\BR(\BXSG)=R\cdot\BR(\BXC)$, and compare the
	results with the CLEO II bound.
	Of the five parameters, $m_t$ is restricted by the CDF bound,
	$m_t^{pole}\simeq174\pm17$~GeV~[23].
	In our analysis, $m_t$ is the running top mass at $M_Z$, which
	is related to the pole mass by~[25]
$$
   m_t^{pole}=m_t\rho^t_z[1+{4\alphasw\over3\pi}+11({\alphasw\over\pi})^2],
\eqno(6)
$$
	where $\rho^t_z\simeq h_t(m_t)/h_t(M_Z)$ is the ratio of the top Yukawa
	couplings at $m_t$ and $M_Z$ [We assume that the Higgs VEVs,
	$v_1$~and~$v_2$, and hence $\tan\beta$, do not change significantly
	between $m_t$ and $M_Z$].
	The pole mass given by eq.~(6) is about $3-5\%$ larger than the running
	mass at $M_Z$.
	The naturalness condition restricts $m_0$ and $m_{\tilde g}$ to be less
	than O(1~TeV), and we allow them to lie between 100~GeV and 2~TeV.
	We parametrize $A_t$ in units of $m_0$, and restrict $|A_t/m_0|\le2.0$.
	Notice also that the top mass is very close to its Landau pole,
	$m_t^{Landau}=C\sin\beta$, where $C\sim195$~GeV.
	This implies that the error in $\alphasw$ plays an important role in
	the RGE analysis of the spectrum.
	Here, instead of varying $\alphasw$, we let $\alpha_G$ vary between
	$1/24.11$ and $1/24.5$ with a fixed $\mgut=10^{16.187}$, since a slight
	change in $\mgut$ does not affect the coupling unification as
	significantly as a change in $\alpha_G$.
	It turns out that $\alpha_G=1/24.11$ gives, at the one-loop level
	and for $m_s=M_Z$, the best fit to $\alphasw=0.118$,
	$\alpha_2(M_Z)=0.03358\pm0.00011$, and
	$\alpha_1(M_Z)\equiv(5/3)\alpha_Y=0.016985\pm0.000020$ for the
	$\mgut$ cited above, and the two-loop corrections are small~[4].
	Decreasing $\alpha_G$ corresponds to decreasing $\alpha_s(M_Z)$.
	We will discuss below the consequences of varying $\alpha_G$.

	We have surveyed a large domain in the parameter space described
	by eq(1). The branching ratio depends importantly on
	the values of the parameters in this 4-dimensional space.
	Significant deviations from the SM value can occur in certain
	regions of the parameter space.
	Characteristically the region where large deviations from SM arise
	occur when $m_0$, $m_{\tilde g}$ are much smaller than their
	naturalness limits and $\tan\beta$ gets large, i.e., typically larger
	than 10.
	Specifically in this region of the  parameter space $\BR(\bsg)$ can be
	significantly below its SM predictions, and for certain points, an
	almost perfect cancellation is observed.
	This is because the chargino-squark penguin diagrams contribute
	destructively to the total amplitude at these points, with a
	coefficient $\sim1/\cos\beta\sim\tan\beta$.
	This destructive interference between various sources in MSSM has also
	been observed previously~[26] [the symmetric distribution of the
	branching ratios around the SM values found in Ref.~[26] is because
	those authors allowed $\tanbeta$ to be as large as~60].
	Similar deviations from the SM value can also occur for smaller
	$\tanbeta$, although this is less frequent.
	For $\tan\beta< 10.0$, the deviations of $\BR(\bsg)$ from the SM
	values are less dramatic.
	In this region, the current CLEO bound is not stringent enough to
	strongly constrain the MSGM.
	However, with a moderate, e.g., about 30\%, improvement in the
	CLEO bound constraints on the model will emerge.

	Figures~1,~2, and~3 show plots of $\BR(\bsg)$ as a function
	of the light chargino masses, $m_{{\tilde W}_1}$, and the soft
	supersymmetry breaking parameter, $m_0$, for $\tan\beta=5.0$,
	$|A_t/m_0|=0.5$, $m_t=160,~170$~GeV, and $\alpha_G^{-1}=24.11,~24.5$.
	In these graphs, all the masses are in units of~GeV.
	The graphs for $m_t=150$~GeV are similar to Figures~1,~2, and~3, except
	that the branching ratios are smaller.
	We also impose a phenomenological lower bound on the light chargino
	mass, i.e., $m_{{\tilde W}_1}>45$~GeV.
	These figures are characterized by four parameters, $m_t$,
	$\alpha_G^{-1}$, and the signs 	of $A_t$ and $\mu$.
	The figures labeled by `a' have $A_t<0$, $\mu>0$, those by `b'
	have $A_t<0$, $\mu<0$, those by `c' have $A_t>0$, $\mu>0$
	and those by `d' have $A_t>0$, $\mu<0$.
	These figures contain the following results.
	(1)
	The $\bsg$ branching ratio at the points at which $A_t$ and $\mu$
	have the same sign is in general larger than that at the points where
	$A_t$ and $\mu$ have the opposite sign.
	This can be seen by comparing Figures~1a and~1d with Figures~1b and~1c.
	This is a generic feature of the MSGM, because eq.~(5) gives a smaller
	light stop mass when $A_t$ and~$\mu$ have the same sign.
	The gaps in the lines for $m_0=1000$~GeV in Figure~1a and for $m_0=600$
	and $1000$~GeV in Figure~1b are due to the light stop turning tachyonic.
	These gaps occurs only for $A_t$ negative for the region in
	parameter space we have studied.
	Generally, the constraint that the light stop remains
	non-tachyonic is most stringent for $A_t<0$.
	(2) The $\bsg$ branching ratio increases when $m_t$ increases.
	The dependence on $m_t$ is correlated with other parameters.
	The parameters for Figure~1 ($m_t=160$, $m_t^{pole}\simeq165$) and those
	for Figures~3 ($m_t=170$, $m_t^{pole}\simeq175$) differ only in $m_t$.
	Remarkable is that a slight change in $m_t$ (about 6\%), for $A_t$
	negative (Figures~1a and~1b vs. Figures~3a and~3b) eliminates a large
	part of the allowed parameter space.
	For example, while the values of $m_0$ ranging from 100 to~1000~GeV
	are all allowed for $m_t=160$~GeV, for both positive (1a) and negative
	(1b) $\mu$, the only allowed values of $m_0$ are below 200~GeV for
	$m_t=170$~GeV for both positive (3a) and negative (3b) $\mu$.
	However, if $A_t$ is positive (Figures~1c, 1d,~3c,~and~3d),
	the same change in $m_t$ does not significantly affect the allowed
	parameter domain.
	The reduction in parameter space can again be explained by the light
	stop turning tachyonic.
	(3) Most interesting is the $\alpha_G$ dependence of the branching
	ratio.
	To see this effect, let us compare Figures~1 [$\alpha_G^{-1}=24.11$,
	for which $\alpha_s(M_Z)=0.118$] with Figures~2 [$\alpha_G^{-1}=24.5$,
	for which $\alpha_s(M_Z)=0.113$].
	For $A_t$ negative, we again  find that a larger part of the parameter
	space are excluded (1a~vs.~2a, and 1b~vs.~2b), while for $A_t$ positive,
	the allowed parameter space remains almost the same (1c~vs.~2c, and
	1d~vs.~2d).
	Although the light stop turning tachyonic is the reason for this,
	the physics involved is quite different from the above.
	Qualitatively, one can attribute this phenomenon to the fact that~$h_t$
	is very close to its Landau pole.
	Namely, a reduction in $\alpha_G$ modifies various form factors defined
	in Ref.~[24], making $h_t$ closer to its Landau pole.
	This in turn is reflected in the light stop turning tachyonic.
	The combined effects of simultaneous change in $m_t$ and $\alpha_G$
	is very dramatic --- the only allowed parameter space is $m_0=100$~GeV
	for $m_t=170$ and $\alpha_G^{-1}=24.5$ when $A_t$  is negative!
	This is because the simultaneous changes in both $m_t$ and $\alpha_G$
	add up almost multiplicatively to aggravate the closeness to the
	Landau pole~[4].
	For $A_t$ positive, the change is not as large, and the allowed
	domain is still large.
	Similar reduction of the allowed domain also exists for $A_t$ with
	other negative values -- the more negative $A_t$ is, the smaller
	the allowed domain remains.

	In conclusion, we have performed a detailed study of the constraints
	from the $\bsg$ decay on the MSGM.
	There are regions of the parameter space where the branching ratio
	exceeds the CLEO II bound, and this region is excluded .
	However, there still exists a large domain that satisfies the CLEO
	bound.
	A more accurate determination of the branching ratio would further
	constrain this model.
	An interesting result of the analysis is that very little allowed
	domain of the parameter space was found for $A_t<0$ and $m_t$ in
	vicinity of the CDF central value of 174~GeV for $A_t<0$.
	Thus the allowed domain resides mostly in $A_t>0$ region.

	This work was supported in part under NSF Grant Nos. PHY-916593 and
	PHY-9306906.
\vfill
\eject
\noindent{\bf References}
\def\wss#1{World Scientific, Singapore, 19#1}
\def\etal{{\it et.~al.}}
\def\pran{P.~Nath}
\def\dick{R.~Arnowitt}
\def\jizhi{J.~Wu}
\item{[1]} P.~Langacker, Proceedings, PASCOS 90 Symposium, Editors, \pran,
	and S.Reucroft (\wss{90});
	J.~Ellis, S.~Kelly, and D.~V.~Nanopoulos, \PLB{249}{441}{90},
	\ibid{\bf260B}{131}{91};
	U.~Amaldi, W~de~Boer, and H.~Furstenau, \ibid{\bf260B}{447}{91}.
\item{[2]} For reviews, see: \pran, \dick, and A.~H.~Chamsedine, "Applied
	N=1 Supergravity", (\wss{84});
	H.~Haber, and G.~Kane, \PRT{117}{75}{85}.
\item{[3]} For a detailed discussion, see: \dick, and \pran, Lectures at
	VII J.~A.~Swieca Summer School, Campos de Jordao, Brazil, 1993
	(\wss{94}).
\item{[4]} \pran, \jizhi, and \dick, in preparation.
\item{[5]} \dick, and \pran, \PRD{49}{1479}{94}.
\item{[6]} \dick, and \pran, \PLB{299}{58}{93}, and Erratum,
	\ibid{\bf303B}{403}{93};
	\pran, and \dick, \PRL{70}{3696}{93}.
\item{[7]} \pran, and \dick, CERN-TH.7214/94/NUB-TH./CTP-TAMU-32/94.
\item{[8]} R.~Ammar \etal (CLEO Collaboration), \PRL{71}{674}{93}.
\item{[9]} N.~G.~Deshpande, P.~Lo, J.~Trampetic, G.~Eilam, and P.~Singer,
	\PRL{59}{183}{87};
	T.~Altomari, \PRD{37}{677}{87};
	C.~A.~Dominguiz, N.~Paver, and Riazuddin, \PLB{214}{459}{88}.
\item{[10]} R.~Casalbouni, A.~Deandrea, N.~Di.~Bartolomeo, R.~Catto, and
	G.~Nardulli, \PLB{312}{315}{93}.
\item{[11]} P.~Colangelo, C.~A.~Dominguiz, G.~Nardulli, and N.~Paver,
	\PLB{317}{183}{93}.
\item{[12]} A.~F.~Falk, M.~Luke, and M.~J.~Savage, \PRD{49}{3367}{94}.
\item{[13]} R.~R\"uckl, Max-Planck-Institute Report, MPI-Ph/36/89 (1989).
\item{[14]} B.~Grinstein, R.~Springer, and M.~B.~Wise, \NPB{339}{269}{89};
	M.~Misiak, \PLB{269}{161}{91}.
\item{[15]} P.~Cho, and M.~Misiak, Caltech Report, CALT-68-1893 (1993).
\item{[16]} M.~Misiak, \NPB{393}{23}{93};
	K.~Adel, and Y.~P.~Yao, \MPLA{8}{1679}{93}, and UM-TH-93-20 (1993);
	M.~Ciuchini, E.~Franco, G.~Martinelli, L.~Reina, and L.~Silverstini,
	\PLB{316}{127}{93}.
\item{[17]} M.~Misiak, \PLB{321}{113}{94}.
\item{[18]} A.~J.~Buras, M.~Misiak, M.M\"unz, and S.~Porkorski,
	Max-Planck-Institute Report, MPI-Ph/93-77 (1993).
\item{[19]} S.~Bertolini, F.~Borsumati, A.~Masiero, and G.~Ridolfi,
	\NPB{353}{541}{93}.
\item{[20]} J.~Hewett, \PRL{70}{1045}{93};
	V.~Barger, M.~Berger, and R.~J.~N.~Phillips, \PRL{70}{1368}{93};
	M.~A.~Diaz, \PLB{304}{2}{93}.
\item{[21]} R.~Barbieri, and G.~Guidice, \PLB{309}{86}{93}.
\item{[22]} J.~L.~Lopez, D.~V.~Nanopoulos, and G.~Park, \PRD{48}{R974}{93};
	Y.~Okada, \PLB{315}{119}{93};
	R.~Garisto, and J.~N.~Ng, \PLB{315}{372}{93};
	S.~Bertolini, and F.~Vissani, SISSA 40/94/EP (1994).
\item{[23]} CDF Collaboration, Fermilab-Pub-94/097-E (1994).
\item{[24]} L.~E.~Iba\~nez, C.~Lopez, and C.~Mu\~nos, \NPB{256}{218}{85}.

\def\ZPC#1#2#3{Z. Phys. {\bf C} #1, #2 (19#3)}

\item{[25]} H.~Arason, D.~J.~Casta\~no, B.~Kesthelyi, S.~Mikaelian,
	E.~J.~Piard, P.~Ramond, and B.~D.~Wright, \PRD{46}{3945}{92};
	N.~Gray, D.~J.~Broadhurst, W.~Grafe, and K.~Schilcher,
	\ZPC{48}{673}{90};
	R.~Hampfling, DESY Report, DESY94-057 (1994).
\item{[26]} G.~Kane, C.~Kolda, L.~Roszkowski, and D.~J.~Wells, University of
	Michigan Report, UM-TH-93-24 (1993).

\vfill
\eject
\medskip
\medskip
\noindent{\bf Figure Captions}
\item{Fig.~1} The branching ratio, $\BR(\bsg)$, plotted with respect to the
	light chargino mass, $m_{{\tilde W}_1}>45$ GeV, for $\tan\beta=5.0$,
	the running mass $m_t=160$~GeV, $|A_t/m_0|=0.5$,
	and $\alpha_G^{-1}=24.11$, and for
	(a) $A_t<0$, $\mu>0$; (b) $A_t<0$, $\mu<0$; (c) $A_t>0$, $\mu>0$;
	and (d) $A_t>0$, $\mu<0$.
	All masses are in units of GeV.
	The standard model gives a branching ratio,
	$\BR(\bsg)=3.55\times10^{-4}$.
	The discontinuity in the $m_0=1000$~GeV line for~(a) and in the
	$m_0=600$, 800~and 1000~GeV lines for~(b) is due to the light stop
	turning tachyonic.
\item{Fig.~2} Same as Fig.~1 for $\alpha_G=1/24.5$.
	The discontinuity in the $m_0=1000$~GeV line for~(a) and for~(b) is
	also due to the light stop turning tachyonic, as in Figure 1.
\item{Fig.~3} Same as Fig.~1 for the running mass $m_t=170$~GeV.
	The standard model gives a branching ratio,
	$\BR(\bsg)=3.66\times10^{-4}$.

\vfill
\eject
\end
\bye